\documentstyle[12pt]{article}
\def\theequation{\arabic{section}.\arabic{equation}}
\newcommand{\beq}{\begin{equation}}
\newcommand{\eeq}{\end{equation}}

\newcommand{\dd}{D\hspace{-.65em}/}

\newcommand{\rf}[1]{(\ref{#1})}
\newcommand{\eq}[1]{eq.~(\ref{#1})}

\def\dop{Dirac operator}

\begin{document}
%\begin{flushright}
%HUTP-95/A005\\
%NBI-HE-95-02 \\
%January 1995\\
%\end{flushright}
%\vspace{0.5cm}
\begin{center}
{\large {\bf A ${\bf Z_2}$ Structure in the Configuration Space
of Yang-Mills Theories}}\\
\vspace{1.5cm}
{\bf Minos Axenides}
\footnote{present address: Dept.of Physics,
University of Crete, GR-71409 Iraklion-Crete,\\ Greece.
e-mail:axenides@nbivax.nbi.dk}\\
\vspace{0.4cm}
{\em The Niels Bohr Institute\\
University of Copenhagen, 17 Blegdamsvej, 2100 Copenhagen, Denmark}\\
\vspace{0.4cm}
{\bf Andrei Johansen}
\footnote{e-mail:johansen@string.harvard.edu /
ajohansen@nbivax.nbi.dk}\\
\vspace{0.4cm}
{\em The St. Petersburg Nuclear Physics Institute\\
Gatchina, St.Petersburg District, 188350 Russia}\\
\vspace{0.4cm}
and\\
\vspace{0.4cm}
{\em Lyman Laboratory of Physics, Harvard University\\
Cambridge, MA 02138 USA}\\
\vspace{0.4cm}
{\bf Jesper M\o ller}
\footnote{e-mail:moller@math.ku.dk}\\
\vspace{0.4cm}
{\em Mathematical Institute\\
Univ. of Copenhagen, Universitetsparken 5, 2100 Copenhagen,
Denmark}\\
\vspace{0.4cm}
\end{center}

%\baselineskip=2\bskip
\begin{abstract}
We argue for the presence of a ${\bf Z}_2$ topological structure in
the space
of static gauge-Higgs field configurations of $SU(2n)$ and $SO(2n)$
Yang-Mills
theories.
We rigorously prove the existence of a ${\bf Z}_2$ homotopy group of
mappings from the 2-dim. projective sphere ${\bf R}P^2$ into
$SU(2n)/{\bf Z}_2$ and $SO(2n)/{\bf Z}_2$ Lie groups respectively.
Consequently the symmetric phase of these theories admits infinite
surfaces
of odd-parity
static and unstable gauge field configurations which divide
into two
disconnected sectors with integer Chern-Simons numbers $n$ and
$n+1/2$
respectively.
Such a ${\bf Z}_2$ structure persists in the Higgs phase of the above
theories and accounts for the existence of
 $CS=1/2$ odd-parity saddle point solutions to the field
equations
which correspond to spontaneous symmetry breaking mass
scales.

\end{abstract}

\newpage
\section{Introduction}
\setcounter{equation}{0}

The electroweak interactions of
quarks and leptons are well described by a gauge theory of
an $SU_{I}(2) \times U_{Y}(1)$ of isospin ($I$) and
hypercharge ($Y$) spontaneously broken to $U_{em}(1)$ of
electromagnetism.
It is a theory of the Yang-Mills  $SU(N)$ type.
As a consequence the vacuum has a nontrivial periodic
structure \cite{Jackiv}.
The distinct ground
states are labeled by integer values of the Chern-Simons number (CS)
of the three dimensional (spatial component)
$SU(N)$ gauge field $W_i ,$ with the index $i=1,2,3$ corresponding to
spatial
directions in the four dimensional Minkowski space.
The Chern-Simons number is given by the following functional
\beq
CS [W] = \frac{1}{16\pi^2} \int_{D^3}
{\rm Tr} (WdW -\frac{2i}{3} W^3)
\label{cs}
\eeq
where $D^3$ is a 3-dimensional disk.
This structure follows from the topological property that
$\pi_{3}(SU(N))=Z$ \cite{Jackiv}.

The height of the potential barrier between the adjacent vacua is
arbitrary for an unbroken $SU(N)$ gauge theory.
Quantum tunneling is
induced by instantons, finite action solutions to the 4-d Euclidean
equations of motion. Large (small) size instantons transverse low
(high)
barriers respectively. In the case of a spontaneously broken $SU(N)$
theory the height of the barrier is fixed and is roughly proportional
to
the mass scale of the theory $M_w /\alpha$  ($\alpha =$ gauge
coupling
constant, and $M_w$ is a mass of the vector bosons).
Quantum tunneling through instantons is exponentially
suppressed by their finite action \cite{tHooft}.
 More specifically for a chiral
theory such as the electroweak Weinberg-Salam model Baryon number is
violated through the chiral anomaly. A change in Chern-Simons number
through transitions between different vacua results in a change of
the net Baryon number.

While at zero temperature B-violating transitions are
exponentially suppressed at sufficiently high temperature
$T\leq M_{w}/\alpha$ it has been argued \cite{cohen} that they become
quite rapid as they are dominated by a static, finite energy unstable
solution to the electroweak equations of motion, the sphaleron
\cite{Klink}.
This is
because at precisely this temperature range  $M_{w}<T<M_{w}/\alpha$
the sphaleron configuration is a saddle point on the energy surface,
located
at the highest point of a continuous set of static configurations
that
interpolates between the topologically distinct vacua with
$CS=n,n+1(n\in {\bf Z})$ and has $CS=n+1/2$.
In thermal equilibrium and at temperatures
$T\leq E_{sp}=M_{w}/\alpha$ the probability of forming a coherent
sphaleron in the hot plasma is given by the Boltzmann weight of the
classical sphaleron energy $P\sim \exp (- \frac{E_{sp}}{T})$
\cite{peter}.

Though the sphaleron configuration in the standard electroweak
theory is homotopically trivial
because of $\pi_2 (SU(2))=0$, the existence of saddle point
solutions in spontaneously broken gauge
theories is a consequence of a nontrivial topological
structure in their configuration space.
(Configuration space is hereby
taken to be the space of all static, finite energy
$3$-dim. configurations of Yang-Mills-Higgs fields $W_i (\vec{x}),\;
\; \phi (\vec{x}).$)
This, in turn,
is a reflection of their nontrivial periodic vacuum structure.
It was in fact rigorously argued
that saddle point solutions must be a consequence
of the existence of noncontractible loops in the bosonic sector of
$SU(2)$ Yang-Mills theories \cite{Manton} such as in the classical
Weinberg-Salam model.
While the existence of such a noncontractible loop
along with a sphaleron saddle point is a symptom of the nontrivial
topology of the $SU(2)$ configuration space yet by themselves they
tell
us nothing more about what is its actual character.
A crucial condition for the existence of
saddle point solutions with  $CS=1/2$  is the
presence of a mass scale, or equivalently of a mass gap in the
subspace
of configurations with  $CS=n+1/2$  which is introduced through the
Higgs mechanism.
This very condition is somewhat restrictive and not a
necessary ingredient of the nontrivial structure we are after as we
will
argue shortly.

Indeed in a nonabelian pure  $SU(2)$  gauge theory with no such mass
scale
$(M_{w}\to 0),$  and in the absence of a saddle point sphaleron
solution it is unclear which finite energy static
configurations with $CS=1/2$ characterize
the nontrivial topological structure of its configuration space.
The latter is of course expected to be present and
reflect the existence of a periodic vacuum.
In the context of the electroweak theory it is unknown
which are the sphaleron-like configurations which
lie deep in its high temperature symmetric phase and
mediate unsuppressed baryon violating thermal transitions.
Moreover the very origin of the B-violating single normalizable
fermionic zero
mode
in the presence of the $CS=1/2$ electroweak sphaleron has also been
obscure.
We have recently addressed both of these issues in
the context of the  $SU(2)\times U(1)$ Weinberg-Salam model
\cite{Topclass}.
There we observed that the electroweak sphaleron written
in an appropriate gauge is odd under
parity reflection symmetry with an odd pure gauge behavior at
infinity.
This was sufficient for it to possess a  $CS=1/2$  and a single
fermionic zero mode.
Moreover we showed that the number of fermionic zero modes
modulo $2$ in the background of gauge and
Higgs static configurations odd under a
properly defined parity is a  topological invariant which is
determined by the parity properties of the  pure gauge behavior
at spatial infinity.
Configurations with even pure gauge behavior
possess a $CS=n$ ($n$ is an integer) and an even number of fermionic
zero
modes.
The ones
with odd pure gauge behavior at infinity have a $CS=n+1/2$ and an odd
number of fermion zero modes mediating as a consequence fermionic
level
crossing.
We pointed out that the above
classification reflects the existence of nontrivial
discrete groups of maps $[{\bf R}P^2, SU(2)/{\bf Z}_2]={\bf Z}_2$
as well as $[{\bf R}P^3,SU(2)/{\bf Z}_2]=\hat{ {\bf Z}} $ ($\hat{
{\bf Z}} $
is a double cover of ${\bf Z}$) in
the configuration space of gauge fields (for a strict mathematical
treatment of these statements see next section).

In our present study we will provide rigorous proofs for the
existence of such nontrivial groups of mappings.
We will do it for
the most general case of $SU(N)$ and $SO(N)$ Lie groups.
For the
case of  $N=even,\;\; (N=2n)$ we will establish the existence of a
nontrivial two element $({\bf Z}_2)$ group of maps from the
2-dimensional
projective sphere ${\bf R}P^2= S^2/{\bf Z}_2$  at spatial infinity
into the $SU(2n)/{\bf Z}_2$ and $SO(2n)/{\bf Z}_2$
groups respectively.
More schematically we will demonstrate that
$[{\bf R}P^2,SU(2n)/{\bf Z}_2] = [{\bf R}P^2,SO(2n)/{\bf Z}_2] =
{\bf Z}_2$.
Moreover we will argue that the presence of a $U(1)$
factor in a product group of the type $SU(2n) \times U(1)$  preserves
such a ${\bf Z}_2$ structure.
As our arguments are based on abstract
topological properties of Lie groups they are completely general.
More specifically for the case of the electroweak theory they
demonstrate rigorously the existence of a ${\bf Z}_2$ structure in
its configuration space of static, finite energy gauge fields alone
which are odd under a properly defined parity.
Such a structure provides us with
a novel top-bottom argument for the existence of sphalerons and their
nontrivial
deformations such as the W and Z vortex loops.
As such it does not depend on
the existence of any mass scale in the theory it characterizes
its complete configuration space.
Our arguments identify $SU(2n)$, $SO(2n)$ and $E_7$
as the gauge groups
which admit an similar ${\bf Z}_2$ homotopy group of mappings.

The paper is organized as follows:\\
in section 2 we summarize our
physical arguments for the existence of a ${\bf Z}_2$ structure in a
pure $SU(2)$ Yang-Mills theory.
They mean to physically motivate our
subsequent rigorous mathematical treatment.
In section 3 we prove that for $SU(2n)$ groups
$[{\bf R}P^2,SU(2n)/{\bf Z}_2] =
{\bf Z}_2 $ and that $[{\bf R}P^3, SU(2n)/{\bf Z}_2]= \hat{ {\bf Z}}
$,
where ${\bf \hat{Z}}$ in the r.h.s. of the latter acts as a double
covering of ${\bf Z}$ on the space of odd-parity gauge field
configurations.
Such a properly understood double covered ${\bf Z}$ group can be
interpreted from a physical point of view as ${\bf Z} \times {\bf
Z}_2$
which agrees with the statement of ref.\cite{Topclass}.
In
section 4 we show that the presence of a $U(1)$ factor group does not
affect the ${\bf Z}_2$ structure.  We also show that $[{\bf R}P^2,
SO(2n)/{\bf Z}_2] = {\bf Z}_2$.  We close with some final comments
with regard to the physical implications of the presence of a ${\bf
Z}_2$ topological structure in the configuration space of $SU(2n)$
and $SO(2n)$ gauge theories.
In Appendix
we illustrate a difference between $SU(even)$ from $SU(odd)$ gauge
groups by proving the existence of
a 3-element homotopy group ${\bf Z}_3$ of maps from
$S^3/{\bf Z}_3$ to $SU(3)/{\bf Z}_3$
(more precisely $[S^3/{\bf Z}_3, SU(3)/{\bf Z}_3]= {\bf Z} \oplus
{\bf Z}_3$).
In contrast to the $SU(2n) $ gauge theories it implies a ${\bf Z}_3$
``parity''
classification on the space of 4-dimensional gauge field
configurations
of an $SU(3)$ gauge theory.

\section{${\bf Z}_2$ structure of an $SU(2)$ Configuration Space}
\setcounter{equation}{0}

We start with the observation that our familiar static
sphaleron configuration has an odd-parity gauge field everywhere in
space.
By imposing the same property on all its possible deformations
(they may not be solutions) we find two topologically distinct
sectors of configurations that depend on the (even-odd) parity
properties of their pure gauge behaviour at spatial infinity.

We hereby give a summary of our arguments.
The Chern-Simons number (CS) for the $SU(2)$ weak sphaleron is
defined as the functional $CS(W')$ which is given by
\beq
{\rm CS} (W') ={\rm CS} (W) + S_{WZW} (U') ,
\eeq
where $W'$ is given by
\beq
W'_k = U' W_k (U')^{-1} +i \partial_k U' \; (U')^{-1} ,\;\;\;
k=1,2,3.
\eeq
and $S_{WZW}$ is the Wess-Zumino-Witten functional
\beq
S_{WZW}(U') = \frac{1}{24\pi^2} \int_{D^3} {\rm Tr}
(dU'\; U'^{-1})^3   .
\eeq
Here $W$ is the the sphaleron gauge field
\beq
W_k = f(r) \frac{\epsilon_{ijk} x^i \tau^j}{r^2}
= -if(r) \partial_k U_{sph} \; U_{sph}^{-1} ,
\eeq
where $k =1,2,3,$ $r=|x|,$ and
\beq
U_{sph} =i\frac{(\vec{\tau}, \vec{x})}{r}.
\eeq
We take $U'$ to be an $SU(2)$ group element which is smooth
everywhere and coincides with $U_{sph}$ at spatial infinity.
It is easy to see that $CS (W)=0$ due to the oddness of $W$ under
parity reflections.
Therefore the behavior of $U'$ at the boundary $S^2$ of a $D^3$ disk
determines the value of $CS (W')$ and makes it equal to $1/2 + n$
($n$ is an integer).
For the particular choice of
\beq
U' = \exp
\frac{i\pi}{2} \frac{(\vec{\tau}, \vec{x})}{\sqrt{x^2 + \rho^2}}
\eeq
it gives $S_{WZW}(U')=1/2.$
There are two
ingredients necessary for the result $CS(W')=1/2$.  Firstly from
eq.(2.1) it is necessary that  $CS(W)=0$  which the case as the field
$W$ is odd under parity, i.e. $W(-x)=-W(x).$
Secondly it is the
oddity of $U'$ under parity  $(U'(-x)=-U'(x))$ that renders
$S_{WZW}(U')=1/2$ in eq.(2.3).
Indeed for the choice of an even
$U'$ under parity it is that  $S_{WZW}(U')=0$.

In fact we have argued \cite{Topclass} that for any  $SU(2)$  gauge
field
with pure gauge behavior on a $S^2$  sphere at infinity
\beq
A_i = -i(\partial_i U)\; U^{-1},
\eeq
there exists a nontrivial ${\bf Z}_2$ homotopic classification in the
space of 3-dim. odd-parity gauge fields.
It should be noted that the statement of oddity of a gauge field
configuration under parity reflections is certainly
not a gauge invariant one.
For our arguments to be meaningful though it
suffices that there exists a gauge in which this is true.
As such among
the spatially odd gauge fields the ones with odd $U$ fields
have $CS=1/2$ and the ones with even $U$ fields possess $CS=0$.
It was pointed out that this classification is a
consequence of the
existence of a nontrivial homotopy
group of maps from the projective sphere ${\bf R}P^2 =
S^2/{\bf Z}_2$ (where ${\bf Z}_2$ is a group of parity reflections
with respect to some point in 3-dim. space) to the group $SO(3) =
SU(2)/{\bf Z}_2$ (where ${\bf Z}_2$ is the center of $SU(2)$).
In short $[{\bf R}P^2, SO(3)] = {\bf Z}_2$.
Consequently there exists
odd parity $SU(2)$ gauge fields which split into two topologically
disconnected classes.
It is not possible to get from one to the
other continuously through odd-parity gauge field configurations.

In fact configurations with even $U$ fields are continuously
connected to the
vacua $A_i^{n}= i(\partial_i U_n)
U_n^{-1}$ where the group element $U_n$ is given by the even-parity
(at infinity) group elements \cite{Ryder}
\beq
U_n = \exp (in\pi
\frac{(\vec{\tau}, \vec{x})}{\sqrt{x^2 + \rho^2}}) ,
\eeq
with $\rho$ being a constant parameter and $n$ an integer.
Vacuum configurations are associated with the different
Chern-Simons numbers given by $n.$
Such a classification is a manifestation of the non-triviality of the
homotopic group  $\pi_3 (SU(2)) = {\bf Z}.$
The group element $U_n$
is a constant matrix at infinity and hence it corresponds to a
compactification of $D_3$ into  $S^3.$
In turn by taking into
consideration the group elements which are odd under parity at
infinity we
compactify $D^3$ into ${\bf R}P^3.$
Thus the relevant homotopy group
is in this case $[{\bf R}P^3, SO(3)] = {\bf Z} \times{\bf Z}_2$
It is worth noticing that this statement seems to be different from
that which appears in the strict mathematical framework in next
section where this homotopy group is shown to be ${\bf Z} .$
However this ${\bf Z}$ group is shown there to be double covered and
can be understood as given above on the physical grounds.

The appearance of an above ${\bf Z}$ factor in the homotopy group can
be better understood in the following way.
The Chern-Simons number of a gauge field configuration is defined as
the
gauge-independent difference in the value of the Chern-Simons
functional of a given gauge field relative to the one of the vacuum.
In order to make this comparison,  we must transform to a gauge in
which the
gauge fields decrease more rapidly than $1/|x|$ at infinity (similar
to the case of sphaleron).
Actually such a gauge transformation is defined modulo a $large$
gauge
transformation which changes the Chern-Simons number by an integer.
The above ${\bf Z}$ factor is actually a manifestation of such an
ambiguity.
It is worth emphasizing that the fractional part of the Chern-Simons
functional is gauge invariant.

A common feature of all odd-parity, even-$U$ configurations is that
they have an integer valued Chern-Simons functional.
Indeed
similarly with the case of the sphaleron we can make a nonsingular
gauge rotation so that we remove the gauge field at infinity.
The
Wess-Zumino-Witten functional would give us a Chern-Simons number for
the gauge field configuration.
It is easy to see that the value of
$S_{WZW}(U)$  is invariant under even-parity smooth deformations of
the $U$ field at infinity.
Indeed a variation of the
Wess-Zumino-Wess functional reads
\beq
\delta S_{WZW} =
\frac{1}{8\pi^2} \int_{D^3} d {\rm Tr} ((U^{-1} \delta U)(U^{-1}dU)^2
).
\eeq
Since the variation of the group element on the surface
$S^2$ is odd under parity and its value depends only on the values of
the fields at the boundary we immediately conclude that the present
variation of the Wess-Zumino-Wess functional equals zero.
On the
other hand let us consider a product of even-parity (at infinity)
group elements $U_1$ and $U_2$ that correspond to any two such gauge
fields.
We have
\beq S_{WZW} (U_1 U_2) = S_{WZW} (U_1) +S_{WZW}
(U_2) + \frac{1}{8\pi^2} \int_{D^3} {\rm Tr} d((U^{-1}_1 dU_1 ) (dU_2
\; U_2^{-1})).
\eeq
The third term in the left hand side of this
equation equals zero due to the odd parity of the integrand at
infinity.
Hence we see that the Wess-Zumino-Witten functional acts
as a homomorphism from the group of maps $U$ to a discrete subgroup
of the group of real numbers which is obviously isomorphic to ${\bf
Z} .$
As we argued before the even-$U$ (at infinity) group element is
contractible to the vacuum.
In turn as it is well known that the
vacuum can have any integer value of the Chern-Simons number we
conclude that even-parity $U$-fields are indeed classified by ${\bf
Z}.$
Thus all odd-parity even-$U$ gauge fields split into an infinite
set of disconnected equivalence classes which are labeled by integer
values of their Chern-Simons numbers.

Let us now consider odd-parity odd-$U$ gauge fields.
A similar argument shows that the value of the Chern-Simons
functional
is a topological invariant while the Wess-Zumino-Witten functional
maps the odd-parity $U$ fields to a discrete subgroup of the group of
real numbers according to eq.(2.10).
On the other hand a product of two odd-parity group elements
$U_1$ and $U_2$ is even under parity.
By taking also into account that the sphaleron has
$S_{WZW} (U) =1/2$ we conclude
that the odd-parity odd-$U$ gauge fields have half-integer
values of the Wess-Zumino-Wess functional and hence are classified
by  $n+1/2$ ($n\in {\bf Z}$)
while these equivalence classes are themselves topologically
disconnected one from the other for different values of $n.$

Thus we see that the Chern-Simons functional plays the role of
a topological charge: it takes values in ${\bf Z}$
for even-$U$ and in ${\bf Z} + 1/2$ for odd-$U$ fields respectively.

An immediate implication of such a topological index for
the fermionic spectrum of a 3-dimensional \dop \ is the following.
Let us consider a \dop \ $\dd = \gamma_i
(\partial_i -iA_i)$ in an external odd-parity gauge field $A_i .$
Its non-zero eigenvalues are paired up ($\lambda, -\lambda$).
Hence when the external field varies continuously the number of
zero modes of the \dop \ is invariant modulo 2.
For the sphaleron background this topological invariant is equal to
one
while for the vacuum its value is zero.
This means that it is not possible to get to the vacuum from
the sphaleron configuration continuously through odd-parity
gauge field configurations.

{}From the above considerations we conclude that in the presence of
an odd-parity external gauge field the number of fermionic zero modes
is 0 mod 2 for even-$U$ and 1 mod 2 for odd-$U$ configurations.

In ref.\cite{Ajhbech} we extended such a classification to the gauge
field configurations which are odd under a
generalized parity reflection symmetry.
This property means an oddity under the parity reflections up to a
gauge
transformation.
We have shown that the ${\bf Z}_2$ structure holds in this case too
and
corresponds to integer and half-integer values of the Chern-Simons
functional.

\section{Compact 1-connected Lie groups}

\setcounter{equation}{0}

We study the topological classification of the maps from ${\bf R}P^2$
and ${\bf R}P^3$ into $G/{\bf Z}_2 ,$ where $G$ is a compact
1-connected
Lie group.
These maps correspond to purely gauge behavior of the gauge
connections
at infinity.

Let $H$ be any compact connected semisimple Lie group, i.e.
\beq
H= G/K ,
\eeq
where $K$ is a subgroup of the center $Z(G)$ of $G.$
Let us consider the associated fibration sequence
\beq
K\to G \to H \to BK\to BG
\label{fibration}
\eeq
$BK$ stands for the classifying space of $K.$
For any pointed space $S$ let $[S,X]$ be the set of pointed homotopy
classes of maps of $S$ into $X.$
If $X$ is a topological, or Lie group, this set is a group.

{\bf Proposition 1}. {\it Let $S$ be any closed 2-dimensional
manifold.
Then}
\beq
\pi_1 :[S,H] \to {\rm Hom} (\pi_1 (S),\pi_1 (H))
\eeq
{\it is a group isomorphism.}

{\em Proof.}
Note that $K =\pi_1 (H) .$
The map $H \to BK$ is a 3-connected $H$-map.
Therefore it induces a bijection
\beq
[S,H] \to [S,BK]
\eeq
of the groups.
Here,
\beq
[S,BK] = H^1 (S;K) = {\rm Hom} (\pi_1 (S),K) =
{\rm Hom} (\pi_1 (S),\pi_1 (H)) . \Box
\eeq

{\bf Corollary 2}.
{\it Let $G$ be any compact simply connected  Lie group
which has a center with ${\bf Z}_2$ as a subgroup.
Then}
\beq [{\bf R}P^2 ,G/{\bf Z}_2 ]\cong {\bf Z}_2 .
\eeq
{\em Proof.}
By proposition 1,
\beq [{\bf R}P^2 ,G/{\bf Z}_2 ]\cong {\rm Hom}
({\bf Z}_2,{\bf Z}_2)={\bf Z}_2.
\eeq
Thus we see that
the space of the gauge field configurations odd under the
parity reflections splits into two homotopically inequivalent
classes which correspond to the ${\bf Z}_2$ group.

We now want to extend such a classification to the maps from ${\bf
R}P^3$ into $H.$

Suppose that $S$ is a 3-dimensional closed manifold.
The functor $[S, *]$ applied to \rf{fibration} yields an exact
sequence
\beq
0\to [S,G] \to [S,H] \to [S,BK] \to 0
\label{exact}
\eeq
of groups.
Here we have used that $[S,K] =0$ and $[S,BG] =0$ since $BG$ is
3-connected.
Note that
$$[S,BK] = H^1 (S,K)$$
as before and that
$$[S,G]=H^3 (S; H_3 (G))$$
since the Postnikov fibration $BG \to K(H_3 (G) ,4)$ which is a
5-connected $H$-map induces a 4-connected $H$-map $G\to K(H_3 (G),3)
.$
Thus \eq{exact} is equivalent to the short exact sequence
\beq
0\to H^3 (S; H_3 (G)) \to [S,H] \to H^1 (S;K) \to 0
\eeq
\label{short}
of groups.

{\em Example.} \\
(a). With $G={\rm Spin} (2n) ,$ $n$ even, and $K={\bf Z}_2 \oplus
{\bf Z}_2 ,$
$$H=G/K = {\rm PSpin} (2n) = SO(2n)/{\bf Z}_2$$
and we obtain the exact sequence
$$0\to {\bf Z} \to [{\bf R}P^3 ,SO(2n)/{\bf Z}_2] \to
{\bf Z}_2 \oplus {\bf Z}_2 \to 0$$
of groups.\\
(b). With $G={\rm Spin} (2n) ,$ $n$ odd, and $K={\bf Z}_4 ,$
$$H= G/K ={\rm PSpin}(2n) =SO(2n)/{\bf Z}_2$$
and we obtain the exact sequence
$$0\to {\bf Z}\to [{\bf R}P^3 ,SO(2n)/{\bf Z}_2]\to {\bf Z}_2\to 0$$
of groups.

We now specialize to the case where $S={\bf R}P^3$ is the real
projective sphere and $K=Z_2$ is of order 2.

{\bf Proposition 3.} {\it The map}
\beq
H_3 : [{\bf R}P^3 ;H] \to {\rm Hom} (H_3 ({\bf R}P^3) ,H_3 (H))
\eeq
{\it is a group isomorphism.}

{\em Proof.} The Serre spectral sequence for the fibration
\beq
G\to H\to {\bf R}P^{\infty} =B{\bf Z}_2,
\eeq
which is part of \eq{fibration} contains the short exact sequence
\beq
0\to H_3 (G) \to H_3 (H) \to H_3 ({\bf Z}_2) \to 0
\label{short1}
\eeq
of abelian groups.
Here $H_3 ({\bf Z}_2 ) = {\bf Z}_2 .$
Moreover the functor $H_3$ induces a commutative diagram

\begin{tabular}{lcccccccc}
$\;$ &$\;$ &$\;$ &$\;$ &$\;$ &$\;$ &$\;$ &$\;$ &$\;$ \\
$0$ & $\to$ & $H^3 ({\bf R}P^3 , H_3 (G))$ & $\to$ & $[{\bf R}P^3
,H]$ &
$\to$ & $H^1 ({\bf R}P^3 ,{\bf Z}_2)$ & $\to$ & $0$\\
$\;$ &$\;$ &$\;$ &$\;$ &$\;$ &$\;$ &$\;$ &$\;$ &$\;$ \\
$\;$ & $\;$ & $\cong \downarrow$ & $\;$ & $\downarrow$ & $\;$ &
$\cong \downarrow$ &
$\;$ & $\;$ \\
$\;$ &$\;$ &$\;$ &$\;$ &$\;$ &$\;$ &$\;$ &$\;$ &$\;$ \\
$0$ & $\to$ & ${\cal H}_G$ &
$\to$ & ${\cal H}_H$
& $\to$ & ${\cal H}_{{\bf Z}_2}$ & $\to$ & $0$\\
$\;$ &$\;$ &$\;$ &$\;$ &$\;$ &$\;$ &$\;$ &$\;$ &$\;$ \\
\end{tabular}
\\
where
$${\cal H}_G ={\rm Hom} (H_3 ({\bf R}P^3 ), H_3 (G)) ,$$
$${\cal H}_H ={\rm Hom} ((H_3 ({\bf R}P^3 ), H_3 (H)) ,$$
$${\cal H}_{{\bf Z}_2} ={\rm Hom} ((H_3 ({\bf R}P^{\infty}),
{\bf Z}_2 ) .$$
The upper line is eq.(3.9) with $S={\bf R}P^3$ and the bottom line
is the result of applying the
exact functor ${\rm Hom} (H_3 ({\bf R}P^3) ,*)$ to
\eq{short1}.
The two  vertical maps are group
isomorphisms; use the Universal Coefficient Theorem for the
left one and simply inspect the right one.
Inspection also
shows that the middle vertical map is a group homomorphism.
Now the  5-lemma \cite{Mac} shows that the middle map is a group
isomorphism.

Assume now that $H$ is {\em simple}, i.e. that $G$ is simple and
simply
connected.

{\bf Lemma 4.}
\beq
H_3 (H) \cong {\bf Z} .
\eeq
{\em Proof.} The Serre spectral sequence for the orientable fibration
\beq
H\to B {\bf Z}_2 \to BG
\eeq
implies that
\beq
H_3 (H;{\bf Z}_2) \cong H_3 (B{\bf Z}_2 ;{\bf Z}_2) \cong
{\bf Z}_2 .
\eeq
The extension \rf{short1} shows that $H_3
(H)$ is isomorphic to ${\bf Z}$ or to ${\bf Z} \oplus {\bf
Z}_2 .$
The latter is impossible since $$H_3 (H) \otimes {\bf
Z}_2 \subseteq H_3 (H;{\bf Z}_2)$$ by the Universal
Coefficient Theorem.
Hence $H_3 (H) \cong {\bf Z} .$

Since also $H_3 ({\bf R}P^3 )={\bf Z}$ (which is a special case of
Lemma
4.), we conclude that double covering homomorphism  $G\to H$ induces
a
commutative diagram of abelian groups

\begin{tabular}{ccccccccc}
$\;\;\;\;$ &$\;$ &$\;$ &$\;$ &$\;$ &$\;$ &$\;$ &$\;$ &$\;$ \\
$\;\;\; 0$
& $\to$ & $[{\bf R}P^3 ,G]$ & $\to$ & $[{\bf R}P^3 ,H]$ & $\to$ &
${\bf
Z}_2 \to 0$\\
$\;\;\;\;$ &$\;$ &$\;$ &$\;$ &$\;$ &$\;$ &$\;$ &$\;$ &$\;$ \\
$\;$ & $\;$ & $\cong \downarrow$ & $\;$ & $\cong \downarrow$ & $\;$
& $||\;\;$ & $\;$ & $\;$ \\
$\;\;\;\;$ &$\;$ &$\;$ &$\;$ &$\;$ &$\;$ &$\;$ &$\;$ &$\;$ \\
$\;\;\; 0$ & $\to$ & ${\bf Z}$ & $\stackrel{\times 2}{\rightarrow}$ &
${\bf Z}$ & $\to$ & ${\bf Z}_2 \to 0$ \\
$\;\;\;\;$ &$\;$ &$\;$ &$\;$ &$\;$ &$\;$ &$\;$ &$\;$ &$\;$ \\
\end{tabular}
\\
with exact rows.
Here we could take $G\to H$ to be $SU(2) \to SO(3)$ or
$SU(2n)\to SU(2n)/{\bf Z}_2,$ or $E_7 \to E_7/{\bf Z}_2 .$

This double covering corresponds exactly to the sets of integer and
half integer values of the Chern-Simons functional as it was
described
in the introduction.

The double covering homomorphism $G\to G/{\bf Z}_2$ induces an
isomorphism
$\pi_i(G)= \pi_i(G/{\bf Z}_2)$  for $i\geq 2 !$
In the case of $SU(2n)$ group $\pi_3
(SU(2n))=\pi_3(SU(2n)/{\bf Z}_2)= {\bf Z}.$
For the case $i=1$ however there exists an exact sequence
$$1\to\pi_1(G)\to\pi_1(G/{\bf Z}_2)\to {\bf Z}_2 \to 1$$
which implies that $\pi_1(G)$ is smaller than $\pi_1(G/{\bf
Z}_2).$

\section{$SU(2n)\times U(1)$ and $SO(2n)$ Lie groups}
\setcounter{equation}{0}

Let us now consider the case of $G=SU(N)\times U(1)$ groups.
Assume that ${\bf Z}_2 \subseteq G$ is a central group of
order 2.  Form $H=G/{\bf Z}_2 .$ We are interested in $[{\bf
R}P^2 ,H].$ There are two cases: $A$ and $B.$

$A.$ Suppose that the composite ${\bf Z}_2 \hookrightarrow G
\stackrel{pr}{\rightarrow} U(1)$ is trivial.
Then ${\bf Z}_2 \subseteq SU(N),$ (this is possible of course only
for even $N=2n$) so
\beq
H=SU(N)/{\bf Z}_2 \times U(1)
\eeq
and hence
\beq
[{\bf R}P^2 ,H] = [{\bf R}P^2 ,SU(N)/{\bf Z}_2] \times
[{\bf R}P^2 ,U(1)] ,
\eeq
where
the first factor is $\cong {\bf Z}_2$ by Corollary 2, and the
second factor $[{\bf R}P^2 ,U(1)]=H^1 ({\bf R}P^2 ,{\bf Z})
=0.$

$B.$ Suppose that the composite ${\bf Z}_2 \hookrightarrow G
\stackrel{pr}{\rightarrow} U(1)$ is nontrivial (i.e. that
${\bf Z}_2
\not\subseteq SU(N)$).
In the diagram

\begin{tabular}{ccccccccc}
$\;\;\;\;$ &$\;$ &$\;$ &$\;$ &$\;$ &$\;$ &$\;$ &$\;$ &$\;$ \\
$\;{\bf Z}_2$ & $\to$ & $\; G$ & $\to$ & $\; H$ & $\to$ &
$B{\bf Z}_2$ & $\to$ & $BG$\\
$\;\;\;\;$ & $\;$ & $\;$ & $\;$
& $\;$ & $\;$ & $\;$ & $\;$ &$\;$ \\
$\;$ & $\;$ & $\;$ &
$\;$ & $\;$ & $\;\;$ & $\searrow$ & $\;$ &$\downarrow Bpr$
\\ $\;\;\;\;$ &$\;$ &$\;$ &$\;$ &$\;$ &$\;$ &$\;$ &$\;$ &$\;$
\\ $\;\;\;$ & $\;\;$ & $\;\;$ & $\;\;$ & $\;\;$ & $\;\;$ &
$\;\;$ &$\;\;$ &$BU(1)$ \\
$\;\;\;$ &$\;$ &$\;$ &$\;$ &$\;$
&$\;$ &$\;$ &$\;$ &$\;$ \\
\end{tabular}
\\
the upper row is a fibration sequence and the slanted arrow
is essential (i.e. not nullhomotopic).
Apply the functor $[{\bf R}P^2 , *]$ to this diagram and
obtain the commutative diagram

\begin{tabular}{ccccccccc}
$\;\;\;\;$ &$\;$ &$\;$ &$\;$ &$\;$ &$\;$ &$\;$ \\
$[{\bf R}P^2 ,G]$ & $\to$ & $[{\bf R}P^2 , H]$ & $\to$ &
$[{\bf R}P^2 ,B{\bf Z}_2]$ & $\to$ & $[{\bf R}P^2 ,BG]$ \\
$\;\;\;\;$ &$\;$ &$\;$ &$\;$ &$\;$ &$\;$ &$\;$ \\
$\;$ & $\;$ & $\;$ & $\;$ & $\searrow$ &
$\;$ &$\downarrow \cong$ \\
$\;\;\;\;$ &$\;$ &$\;$ &$\;$ &$\;$ &$\;$ &$\;$
\\
$\;\;\;$ &$\;\;$ & $\;\;$ & $\;\;$ & $\;\;$ &$\;\;$ &$[{\bf R}P^2
,BU(1)]$
\\ $\;\;\;$ &$\;$ &$\;$ &$\;$ &$\;$ &$\;$ &$\;$ \\
\end{tabular}
\\
where the upper row is an exact sequence of sets and the
vertical arrow is a bijection for dimensional reasons.
Here,
\beq
[{\bf R}P^2 ,G] =[{\bf R}P^2 ,SU(N)] \times [{\bf R}P^2
,U(1)] =0
\eeq
since $[{\bf R}P^2 ,SU(N)]=0$ by Proposition 1 and also
$[{\bf R}P^2 ,U(1)]=0$ (as explained under $A$),
[${\bf R}P^2 ,B{\bf Z}_2]\cong H^1 ({\bf R}P^2 ,{\bf Z}_2)
\cong {\bf Z}_2 ,$
and
\beq
[{\bf R}P^2 ,BG] \cong [{\bf R}P^2 ,BSU(N)] \times [{\bf
R}P^2 ,BU(1)] \cong
\eeq
$$\cong 0 \times H^2 ({\bf R}P^2 ,{\bf Z}) \cong {\bf Z}_2
.$$
Moreover, the slanted arrow is a bijection (this follows, for
instance, by applying $[{\bf R}P^2 ,*]$ to the fibration
sequence $U(1)\to B{\bf Z}_2 \to BU(1)$).
Then also
\beq
[{\bf R}P^2 ,B{\bf Z}_2] \to [{\bf R}P^2 ,BG]
\eeq
is a bijection.
Hence $[{\bf R}P^2 ,H] =0$ by exactness.

Thus we see that the presence of $U(1)$ group does not affect
the homotopy.

We now proceed to examine the case of $SO(N)$ groups whose center
is ${\bf Z}_2$, i.e. $N=2n$.
The quotient group $H=SO(2n)/{\bf Z}_2$ is the semi-spinor
group $PSO(2n).$
Its fundamental group is given by
\beq
\pi_1 (PSO(2n)) = \left\{ \begin{array}{c}
{\bf Z}_2 \oplus {\bf
Z}_2 , \;\; n \;\;{\rm even} \\
{\bf Z}_4 ,\;\;\;\;\; n \;\; {\rm odd}. \end{array} \right.
\eeq
By Proposition 1,
\beq
[{\bf R}P^2 ,PSO(2n)] = {\rm Hom} (\pi_1 ({\bf R}P^2) ,\pi_1
(PSO(2n)) =
{\rm Hom} ({\bf Z}_2 ,\pi_1 (PSO(2n))) \cong
\eeq
$$\left\{ \begin{array}{c}
{\bf Z}_2 \oplus {\bf
Z}_2 , \;\; n \;\;{\rm even} \\
{\bf Z}_2 ,\;\;\;\;\; n \;\; {\rm odd}. \end{array} \right.
.$$

Let us discuss some physical aspects of the above
classification.
The nontriviality of the above homotopy groups implies the
existence of infinite surfaces of gauge-higgs fields as well as of
gauge field
configurations alone which are homotopically equivalent to the
sphaleron type configurations for the gauge
theories with $SU(2n)$, $SO(2n)$ and $E_7$ gauge symmetries.
Moreover in the presence of such a topologically non-trivial
gauge field
configuration the three dimensional Dirac operator has an odd number
of normalizable zero modes.
This conclusion obviously applies to the phases both with broken and
unbroken gauge symmetry.
Strictly speaking this fact itself  is not sufficient to prove an
existence of
$saddle$ points of the energy functional in the broken (Higgs) phase.
But with a plausible assumption that the height of the potential
barrier
between vacua is non-zero but not infinitely high
we may expect the presence of saddle point $CS=1/2$ sphalerons
along with their equivalent deformations at each symmetry breaking
mass scale above the $SU(2)$ x $U(1)$ Weinberg-Salam one.
In other words in their Higgs phase these theories possess a
hierarchy of
odd-parity saddle point solutions to the field equations
corresponding to a
sequence of spontaneous symmetry breaking mass scales.

\section{Conclusions}

\setcounter{equation}{0}

We rigorously proved the existence of a topological
classification for odd-parity gauge field configurations in pure
$SU(2n) \times U(1)$ and $SO(2n)$
Yang-Mills theories.
This is an automatic consequence of the existence of
cyclic homotopy groups of maps from the 2-dim. and 3-dim.
projective spheres
into the respective Lie groups appropriately modded out by
${\bf Z}_2$.
This is a sufficient condition for the existence of a nontrivial
${\bf Z}_2$
topological structure in the configuration space of a theory with
spontaneously broken $SU(2)\times U(1)$ gauge
symmetry.
Such a structure is
characterized by the already found $CS=1/2$ electroweak sphaleron.
More importantly it survives
in its absence too for the case of an unbroken gauge symmetry
as it reflects the homotopic properties of the space of gauge field
configurations alone.
In this sense it points to the existence of infinite surfaces of
static, finite energy and
unstable
3-dim. configurations in the symmetric high temperature phase of the
standard electroweak theory which are
homotopically equivalent to the saddle point
sphaleron configuration and have $CS = 1/2$.

More generally as the ${\bf Z}_2$ classification applies to all
$SU(2n)$ and $SO(2n)$ groups as well as for the $E_7$ exceptional
gauge group it indicates the existence of
a hierarchy of saddle point sphaleron configurations associated with
higher gauge groups and symmetry breaking scales beyond the one of
the standard model such as in Grand Unified Gauge theories
\cite{Ross}.
It is worth noticing however that such ``sphalerons'' are not
necessarily
responsible for baryon number violation in realistic models.
The point is that $B+L$ can be gauged (for example in the $SO(10)$
model).
While a hierarchy of saddle point B-violating sphaleron
configurations
could have amusing implications for existing scenaria of Baryogenesis
the physical consequences of B preserving ones are probably of
minimal
significance.

\section{Acknowledgments}

We thank H.B.Nielsen and O.Tornkvist for very valuable discussions.
One of us (A.J.) is grateful to NBI for warm hospitality.
This research of A.J. was supported by a NATO grant GRG 930395.
The research of M.A. was partially supported by a E.U. Network Grant
no. CHRX-CT94-0621

\appendix
\section{Appendix}
\appendix{\em Classification by maps of lens spaces}
\renewcommand{\theequation}{A.\arabic{equation}}
\setcounter{equation}{0}

Here we illustrate the difference between $SU(even)$ and $SU(odd)$
gauge groups
by
considering maps from the lens
space $L=S^3/{\bf Z}_3$ into the group $P=PSU(3) =SU(3)/{\bf
Z}_3$ (the adjoint form of $SU(3)$), where ${\bf Z}_3$ is a
center of $SU(3).$
In particular we prove that $[S^3/{\bf Z}_3, SU(3)/{\bf Z}_3]= {\bf
Z} \oplus
{\bf Z}_3$.

Before to proceed we notice that one can not consider
$\tilde{L}=S^2/{\bf Z}_3$
because it
is not a smooth manifold.
Indeed for a 3-fold covering map $S^2\to \tilde{L}$ the equation
$3\chi
(\tilde{L}) =\chi (S^2)
=2$ ($\chi (Y)$ is an Euler characteristic of $Y$) has no integer
solution.
Hence $\tilde{L}$ can not exist. Therefore we consider $L=S^3/{\bf
Z}_3$ which
is a smooth manifold.
We now can proceed to prove our claim.

{\bf Theorem 1.} $[L,P]\cong {\bf Z} \oplus {\bf Z}_3 .$

Before the proof, we recall some facts about the homology of
these spaces.
The cellular chain complex of $L$ shows that
\beq
H_1 (L) \cong {\bf Z}_3 ,\;\; H_2 (L)=0,\;\; H_3 (L) \cong
{\bf Z} .
\eeq
The Serre spectral sequence
\beq
E^2 = H_* ({\bf Z}_3 ,H_* (SU(3))) \longrightarrow H_* (P)
\eeq
for the fibration
\beq
SU(3) \to P\to B{\bf Z}_3
\label{lensone}
\eeq
has a simple structure partly because
\beq
H_i (B{\bf Z}_3) = \left\{ \begin{array}{c} {\bf Z}_3,\;\;
i\;\; {\rm odd}\\
0, \;\; i\;\; {\rm even} \end{array} \right.
\eeq
The spectral sequence shows that
\beq
H_2 (P) =0
\eeq
and that the sequence
\beq
0\to H_3 (SU(3)) \to H_3 (P) \to H_3 (B{\bf Z}_3) \to 0
\label{lenstwo}
\eeq
is exact.
Here $H_3 (SU(3))\cong {\bf Z}$ and $H_3 (B{\bf Z}_3) \cong
{\bf Z}_3$ so $H_3 (P)$ is either isomorphic to ${\bf Z}$ or
${\bf Z} \oplus {\bf Z}_3.$
In fact

{\bf Lemma 2}.
\beq
H_3 (P) \cong {\bf Z} \oplus {\bf Z}_3 .
\eeq
{\em Proof.} According to R.M.Kane \cite{Kane}
\beq
H^* (P,{\bf F}_3)\cong E(V_3) \otimes E(V_1) \otimes {\bf
F}_3 [V_2]/(U^3_2)
\eeq
where the subscript indicates degree.
In particular
\beq
H^3 (P,{\bf F}_3) \cong {\bf F}_3 \oplus {\bf F}_3.
\eeq
Since $H_2(P) =0,$ $H_3 (P) \otimes {\bf F}_3 \cong H_3
(P,{\bf F}_3) \cong H^3 (P,{\bf F}_3)$ by the Universal
Coefficient Formula.
Therefore we must have $H_3 (P)\cong {\bf Z} \oplus {\bf Z}_3
.$

{\bf Proof of Theorem 1.}
Applying the functor $[L, *]$ to fibration \eq{lensone} and
the functor Hom$(H_3 (L), *)$ to the exact sequence
\eq{lenstwo} yields two exact sequences

\begin{tabular}{ccccccccc}
$\;\;\;\;$ &$\;$ &$\;$ &$\;$ &$\;$ &$\;$ &$\;$ &$\;$ &$\;$ \\
$\;\;\; 0$
& $\to$ & $[L,SU(3)]$ & $\to$ & $[L,P]$ &
$\to$ & $[L,B{\bf Z}_3]$& $\to$& $0$\\
$\;\;\;\;$ &$\;$ &$\;$ &$\;$ &$\;$ &$\;$ &$\;$ &$\;$ &$\;$ \\
$\;$ & $\;$ & $H_3 \downarrow$ & $\;$ & $H_3 \downarrow$ &
$\;$ & $H_3 \downarrow$ & $\;$ & $\;$ \\
$\;\;\;\;$ &$\;$ &$\;$ &$\;$ &$\;$ &$\;$ &$\;$ &$\;$ &$\;$ \\
$\;\;\; 0$ & $\to$ & ${\cal F}_{SU(3)}$ &
$\to$ & ${\cal F}_{P}$ & $\to$ &
${\cal F}_{B{\bf Z}_3}$ & $\to$& $0$ \\
$\;\;\;\;$ &$\;$ &$\;$ &$\;$ &$\;$ &$\;$ &$\;$ &$\;$ &$\;$ \\
\end{tabular}

where
\beq
{\cal F}_{SU(3)} = {\rm Hom} (H_3(L),H_3(SU(3))) ,
\eeq
$${\cal F}_P = {\rm Hom} (H_3(L),H_3(P)) ,$$
$${\cal F}_{B{\bf Z}_3} = {\rm Hom} (H_3(L),H_3(B{\bf Z}_3))
,$$
Connected by a homomorphism determined by the functor $H_3.$
The upper sequence is exact since $P\to B{\bf Z}_3$ is
3-connected so that $[L,P] \to [L,B{\bf Z}_3]$ is surjective.
The lower sequence is exact since $H_3 (L) \cong {\bf Z} .$
The left vertical homomorphism is an isomorphism since $L$ is
a compact orientable 3-manifold.
An inspection shows that the right vertical homomorphism is
an isomorphism.
Hence also the middle homomorphism
\beq
H_3: [L,P] \to {\rm Hom} (H_3 (L),H_3 (P))
\eeq
is an isomorphism.
We have
\beq
{\rm Hom} (H_3 (L),H_3 (P)) \cong H_3 (P) \cong {\bf Z}
\oplus {\bf Z}_3
\eeq
by Lemma 2 and since $H_3 (L) \cong {\bf Z}.$

Let us discuss the physical interpretation of the above
conclusion.
When we consider the lens space $L$ we restrict ourselves to
the four dimensional gauge field configurations which are
invariant under ${\bf Z}_3$ transformations up to gauge
transformations.
The above computed homotopy group implies that this subset
of the gauge field configurations has a fine structure
governed by the ${\bf Z}_3$ group.
That means that this subset is splitted into three
disconnected classes.
It is important to note that the classification applies to four
dimensional
gauge field configurations in contrast to the $SU(2)$ case where the
existence
of a $Z_2$ structure was a property of the space of $3-dim$ static
configurations.
Notice that this structure does not literally have anything to do
with
sphalerons because
it corresponds to a classification of the boundary conditions for
4-dimensional
gauge fields.
Instead we have to assign this structure to the vacuum of the
4-dimensional
Yang-Mills theory.
It would be interesting to have an interpretation of this structure
in terms of
the spectrum of the 4-dimensional Dirac operator.

\end{document}